\PassOptionsToPackage{table}{xcolor}
\documentclass[acmsmall,screen,manuscript]{acmart}

\usepackage{xspace}
\usepackage{tcolorbox}
\usepackage{fontawesome5}
\usepackage{enumitem}
\tcbuselibrary{listings,breakable}

\usepackage{booktabs}      
\usepackage{tabularx}
\usepackage{framed}
\usepackage[strict]{changepage}

\usepackage{subcaption}

\definecolor{lineEst}{HTML}{F9D2D2} 
\definecolor{lineRef}{HTML}{FDF0BF} 
\definecolor{lineNew}{HTML}{D8EFCF} 

\AtBeginDocument{%
  }

\begin{document}

\title{Are We Just Applying LLMs? What SE Research Studies, Underexplores, and Could Do Next in the LLM Era}


\author{Zeinadsadat Saghi}
\authornote{Both authors contributed equally to this research.}
\email{saghi@usc.edu}
\affiliation{%
  \institution{University of Southern California}
  \city{Los Angeles}
  \state{California}
  \country{USA}
}

\author{Sadra Sabouri}
\authornotemark[1]
\email{sabourih@usc.edu}
\affiliation{%
  \institution{University of Southern California}
  \city{Los Angeles}
  \state{California}
  \country{USA}
}

\author{Souti Chattopadhyay}
\email{schattop@usc.edu}
\affiliation{%
  \institution{University of Southern California}
  \city{Los Angeles}
  \state{California}
  \country{USA}
}

\begin{abstract}

\textit{Is AI taking over software engineering research?} By 2025, AI appeared in more
than half of the research-track papers at ICSE, FSE, and ASE. Yet ubiquity
alone cannot show whether AI expanded the publication portfolio, displaced
other work, or redirected the intellectual agenda. We investigate this
shift through a census of 4,790 research papers (three SE conferences;
2018--2026) and free-form responses from 244 SE researchers discussing the future of the SE research community (ICSE 2026 FoSE survey).

We found that while AI is taking over, it is not simply replacing non-AI research; it is unevenly reorganizing SE research around its capabilities. AI accounted for $93.1\%$ of publication growth, but expansion in some venues and problem areas concealed contraction in others. The LLM wave also changed the field's tools more than its problem boundaries: $92.6\%$ of LLM-wave AI papers revisit established SE problems, while most new problems arise from the introduction of LLMs in the SE process themselves. Engagement with AI remains predominantly downstream: $95.7\%$ of AI papers apply AI or evaluate and support its use, while only $4.3\%$ interrogate the paradigm or develop alternatives. The FoSE survey reveals a similar imbalance between recognition and response. AI-driven change was the second-most frequent concern about the field's future ($20.2\%$), yet only $7.0\%$ of responses mentioned AI when researchers were asked to identify one change the community should make. Together, our findings reveal a field reorganizing an inherited agenda around AI faster than it is questioning that reorganization. We discuss how SE can
respond by building cumulatively across technological waves, strengthening
evaluation and assurance, developing SE-native foundations, and studying
software engineering in an AI-transformed world.
\end{abstract}




\maketitle

\section{Introduction}

\begin{quote}
\emph{``Software engineering research will become an AI application area
\ldots{} and eventually will disappear.''}

\hfill ---Anonymous SE researcher, ICSE 2026 FoSE survey~\cite{fose2026survey}
\end{quote}

In only a few years, large language models have become a recurring presence
across software engineering research~\cite{fan2023large,hou2024large}. They
generate code and tests, repair programs, and support comprehension
\cite{chen2021codex,hitsuji2024harnessing,ahmed2023better,
geng2024multiintent}; interactive assistants have also entered developers'
workflows~\cite{barke2023grounded,vaithilingam2022expectation,
peng2023impact}. LLMs are no longer confined to one corner of the field.

Yet ubiquity alone does not reveal how AI is changing the field. The same rise
in AI prevalence could result from three distinct shifts: an expanding
publication system that accommodates additional work, the replacement of
non-AI research, or a redistribution of attention across venues and problems.
Although each would increase the visible share of AI research, each represents
a different change in the field's publication portfolio. We distinguish among
them by situating AI papers within the complete research-track record from
which they emerged.

Situating AI papers within the complete publication record reveals the scale
and distribution of the shift, but not its intellectual direction. A
technological wave may change how researchers approach familiar problems
without changing which problems the field recognizes. Tracing problem
continuity still leaves another question: how does the resulting research
engage with the technology? A paper may apply AI to an SE task, develop the
evaluations and infrastructure needed to support its use, or interrogate the
paradigm's assumptions and consequences. The balance among these forms of
engagement matters because generated suggestions may be incorrect or
insecure~\cite{nguyen2022copilot,pearce2022asleep,perry2023insecure}, and
models continue to struggle with realistic repository-level
tasks~\cite{jimenez2024swebench}. A field can adopt a technology extensively
without proportionally expanding the work that evaluates its limits, questions
its direction, or develops alternatives.

Existing reviews illuminate the LLM literature itself, but not these
field-level changes. Reviews of LLMs for software engineering catalogue the
models, datasets, optimization strategies, evaluation methods, and SE tasks
represented in that literature~\cite{fan2023large,hou2024large}. Related work has traced how
LLMs enter other research communities, including the roles they play in HCI
research and the validity and reproducibility concerns they
introduce~\cite{pang2025llmification}. These studies establish the breadth of LLM applications and identify important
technical gaps. By construction, however, they begin with papers that already
use LLMs~\cite{fan2023large,hou2024large,pang2025llmification}. They can explain
what the LLM literature contains, but not how its growth changed the surrounding
field: how much of the publication portfolio it occupies, where non-AI research
expanded or contracted, whether the technological wave changed the problems
receiving attention, or how the balance among application, service, and
interrogation shifted.

To examine these field-level changes, we construct a census of 4,790
research-track papers published at ICSE, FSE, and ASE between 2018 and 2026,
including 1,939 AI papers. The census allows us to compare AI and non-AI
publication trajectories, trace problems across technological waves, and
distinguish how papers engage with AI.

The published record, however, cannot show how researchers themselves
understand the shift. We therefore analyze open-ended responses from 244
participants collected through the ICSE 2026 FoSE community survey~\cite{fose2026survey}. Because the selected
questions did not mention AI, the responses reveal when researchers themselves
introduced it into their accounts of the field's current problems, future
concerns, and desired changes.

Our analysis is guided by four research questions:

\begin{description}
  \item[\textbf{RQ1.}] How has AI changed the scale and distribution of SE
  research?

  \item[\textbf{RQ2.}] Did the LLM wave introduce new SE problems or redirect
  attention among established ones?

  \item[\textbf{RQ3.}] How does SE research engage with AI, and how has this
  changed across technological waves?

  \item[\textbf{RQ4.}] How does the SE community perceive and respond to the AI
  shift?
\end{description}

The resulting account is neither one of wholesale displacement nor one of
simple expansion. By 2025, AI constituted the majority of research-track
output and accounted for $93.1\%$ of the increase in average annual publication
volume. Yet this growth was uneven: expansion in some venues and problem areas
concealed contraction elsewhere, with non-AI output declining in only 12 of the 21
substantive problem areas. The LLM wave changed the distribution of attention
more than the boundaries of the research agenda: $92.6\%$ of LLM-wave AI papers
revisit established SE problems, while nearly all new problems arise from
building, using, or governing LLMs themselves. Within this inherited problem map, $95.7\%$ of AI papers either apply AI to an
SE task or evaluate and support its use, while only $4.3\%$ question the
paradigm's assumptions and direction, or develop alternatives. Work that
reframes the paradigm or develops SE-native foundations constitutes just
$0.5\%$ of the AI corpus. The survey reveals a corresponding gap between
recognition and response: AI-driven change was the second-most frequent concern
about the field's future, yet only $7.0\%$ of proposed changes mentioned AI.

Taken together, our findings reveal a technological revolution without an
equivalent problem revolution. AI is changing the scale and distribution of SE
research and the tools used to pursue SE, while leaving the boundaries of its
problem agenda largely inherited. The resulting portfolio is organized
predominantly around applying AI and supporting its use, with comparatively
little research questioning the paradigm's assumptions, reframing established
problems, or developing SE-native alternatives. 

By situating LLM research
within the complete publication portfolio, tracing problem lineage across
technological waves, and distinguishing how papers engage with AI, we
provide an empirical basis for broadening the field's agenda. Critically, this does not require retreating from LLM research. Instead, the evidence grounds four directions for what SE research could do next: 
cumulatively connect LLM findings to knowledge from earlier technological
waves; develop evaluation and assurance that remains meaningful beyond
individual benchmarks and model releases; use program semantics,
specifications, architectures, and other SE knowledge to construct SE-native foundations; and examine how AI changes development practices, expertise,
responsibility, and software lifecycles. 
The opportunity is to move from following model capabilities toward
deliberately shaping the questions, standards, and foundations that guide
software engineering's engagement with AI.

\subsection{From AI for SE to LLMs for SE}

The intersection of artificial intelligence and software engineering predates
the current LLM wave. Reviews of deep-learning research documented applications
across program synthesis, comprehension, defect prediction, testing, repair,
security, and maintenance~\cite{watson2022systematic,yang2022survey}. These
reviews organized the literature around the models employed, the software
artifacts used for training and evaluation, the SE tasks addressed, and the
methodological pitfalls of applying learning-based techniques. The
deep-learning wave had therefore already distributed AI research across much of
the software lifecycle before foundation models arrived.

Foundation models widened this intersection. Codex demonstrated the capacity of
a code-trained language model to generate functionally correct programs from
docstrings~\cite{chen2021codex}, while subsequent studies examined LLM-enabled
software engineering more broadly. Fan et al.\ surveyed its capabilities and
open problems~\cite{fan2023large}, and Hou et al.\ synthesized 395 studies
spanning 85 SE tasks~\cite{hou2024large}. More specialized reviews examine
natural-language code generation~\cite{zan2023large} and software
testing~\cite{wang2023softwaretesting}. Collectively, this literature describes
which models are used, how they are adapted, which benchmarks and evaluation
measures are common, and where technical limitations remain.

A parallel empirical literature examines what these systems mean for software
development in practice. Studies of GitHub Copilot have investigated the
interaction strategies developers adopt~\cite{barke2023grounded}, the gap
between anticipated and experienced usability~\cite{vaithilingam2022expectation},
the correctness and comprehensibility of generated
suggestions~\cite{nguyen2022copilot}, and effects on task completion
time~\cite{peng2023impact}. Other studies expose limits that conventional
functional benchmarks may overlook. Generated code can contain exploitable
weaknesses~\cite{pearce2022asleep}; access to an AI assistant can affect both
the security of users' solutions and their confidence in those
solutions~\cite{perry2023insecure}; and repository-level issue resolution
requires capabilities beyond isolated code generation
benchmarks~\cite{jimenez2024swebench}. This work motivates more realistic
evaluation and assurance, but its primary object remains the performance,
or use of particular AI systems.

These studies answer an important question: \emph{What does the AI-for-SE
literature contain?} Their unit of analysis, however, is ordinarily a corpus
selected because its papers already use AI or LLMs. Consequently, they do not
observe the surrounding publication portfolio from which that literature
emerged. They cannot determine whether AI growth added to or replaced other
research, whether its growth was distributed uniformly across venues and
problem areas, or whether the transition from deep learning to LLMs changed the
problems receiving attention. Our study therefore retains non-AI research as
part of the denominator and compares successive technological waves within the
same publication system.

\subsection{From LLM adoption to disciplinary change}

Software engineering has a long tradition of examining its own research
record. Glass et al.\ classified SE research by topic, approach, method,
reference discipline, and level of analysis~\cite{glass2002research}.
Subsequent bibliometric work used citation and topic analysis to identify
changes in the field's active and declining themes
\cite{garousi2016systematic}. Bertolino et al.\ focused specifically on ICSE,
FSE, and ASE, classifying the research problems, contribution types, and
validation strategies represented in their proceedings
\cite{bertolino2018software}. These studies demonstrate that publication
records can reveal how a research community allocates attention and what forms
of contribution it rewards.

Our analysis shares this field-level orientation but asks a different question.
Earlier mappings characterize the topics or contribution types visible during a
period; they do not connect changes in that portfolio to successive
technological waves. Topic recurrence also does not distinguish a problem
revisited with a new technique from one that has been substantively reframed,
nor does topic frequency reveal whether a paper applies, services, or
interrogates the technology around which its contribution is organized.

Recent work outside SE has begun to examine LLM adoption as a disciplinary
phenomenon. Pang et al.\ describe the ``LLM-ification'' of CHI by classifying
the domains, roles, contribution types, and validity concerns represented in
153 LLM-related CHI papers~\cite{pang2025llmification}. At a broader scale,
Liang et al.\ estimate the increasing presence of LLM-modified text across more
than one million scientific papers~\cite{liang2025mapping}. Their study concerns
the use of LLMs in scientific writing rather than the direction of the research
agenda, but it demonstrates how rapidly LLMs have entered research practice.
More directly, Hao et al.\ associate the adoption of AI tools in the natural
sciences with greater individual productivity and impact but a contraction in
the collective range of topics studied~\cite{hao2026artificial}. Technological
adoption can therefore expand research production while simultaneously
concentrating its intellectual scope.

We extend this disciplinary perspective to software engineering in four ways.
First, we measure AI papers against the complete research-track output of three
flagship venues rather than against an AI-only corpus. Second, we compare the
deep-learning and LLM waves to separate technological change from changes in
the problem agenda. Third, we trace whether LLM-wave papers revisit, reframe,
or introduce SE problems and distinguish the research postures through which
they engage with AI. Finally, we place these publication patterns alongside the
concerns and priorities expressed by SE researchers. The result is not another
map of LLM applications, but an account of how their rise changed the
composition and direction of the field studying them.

\section{Methodology}
\label{sec:method}

We combine two complementary analyses. First, we conduct a population-level
census of the research tracks of ICSE, FSE, and ASE to characterize changes in
publication volume, problem selection, and engagement with AI. Second, we
analyze responses to the ICSE 2026 Future of Software Engineering (FoSE)
pre-survey to examine whether the patterns visible in the publication record
also arise, unprompted, in researchers' accounts of the field. The survey serves
as qualitative resonance with the census, not as independent validation of it.

\subsection{Publication corpus}

We collected the published research-track records for ICSE, FSE, and ASE from
2018 through 2026 using the proceedings and associated bibliographic metadata.
The initial corpus contained 4,858 records. We excluded 68 entries that were not
research papers: 57 podcast or session summaries, six tutorials, four
front-matter records, and one journal-first presentation. The resulting
analytical corpus contains 4,790 papers.

Coverage is complete for all three venues through 2025. The 2026 corpus contains
ICSE and FSE but not ASE, whose proceedings were unavailable at the time of
collection. We therefore mark 2026 as partial and exclude it from comparisons
that require complete annual publication capacity. The corpus does not include
journals, journal-first papers, companion proceedings, specialized tracks,
co-hosted conferences, or workshops.

\subsection{Identifying and classifying AI papers}

We first applied broad, curated keyword lexicons to titles, abstracts, and
author keywords. The lexicons covered classical and graph-based machine
learning, deep learning, and large or foundation models. This deliberately
inclusive screen produced 2,009 candidate records. We then examined each
candidate against the inclusion rule that AI must be a central method or object
of study, rather than an incidental reference. This adjudication removed 67
false positives and three non-paper records, producing a final population of
1,939 AI papers and 2,851 non-AI papers.

We assigned each AI paper to one mutually exclusive technology family:
\textsc{LLM}, non-LLM \textsc{Deep Learning}, or classical \textsc{Machine
Learning}. When multiple indicators were present, \textsc{LLM} took precedence
over deep learning, which took precedence over classical machine learning. This
technology-family variable is distinct from the temporal waves used in our
comparisons. We define 2018--2022 as the \emph{deep-learning wave} and
2023--2026 as the \emph{LLM wave}. Thus, ``LLM wave'' denotes a period in which
LLMs became dominant, not a claim that every AI paper published during that
period used an LLM.

\subsection{Coding framework}

Our codebook separates the software-engineering problem a paper studies from how
it engages with AI. The unit of coding is one paper, and coding is based
primarily on its title and abstract.

\emph{SE problem portfolio.}
Every eligible paper, AI and non-AI, received one of 22 fine-grained problem
codes and one corresponding macro-area code. We developed the controlled
vocabulary iteratively from conference session names, author-provided keywords,
and recurring problem descriptions in titles and abstracts. We consolidated
synonymous labels, separated categories that combined substantively different
problems, and assigned multi-topic papers according to their primary SE problem
rather than the technique they employed. We then grouped the 22 codes into
eight macro areas using the same evidence and the conceptual relationships among
the underlying problems. The mapping is fixed and deterministic. A residual
\textsc{Other} category retains papers that do not fit a coherent substantive
area; we report it for completeness but do not interpret it as a research area.

\emph{Problem lineage.}
For the 1,318 AI papers published during the LLM wave, we compared the problem
posed by each paper with the measured problem map of the 2018--2022 corpus. We
coded a problem as \textsc{Established} when it was already present in the
earlier wave, \textsc{Reframed} when an inherited problem was substantively
redefined, and \textsc{New} when the problem was absent from the earlier map.
Lineage refers to the problem posed, not to the novelty of the model, technique,
or claimed contribution. We subsequently open-coded the 90 \textsc{New} papers
into five mutually exclusive categories describing the kind of new problem
introduced.

\emph{Research stance.}
We coded all 1,939 AI papers according to the position of their primary
contribution relative to AI. \textsc{Apply} papers use or adapt an AI technique
for an SE task. \textsc{Service} papers evaluate, characterize, improve, or
provide infrastructure for its use. \textsc{Interrogate} papers question the
direction or assumptions of the prevailing paradigm, empirically test it against
SE-derived requirements, or develop SE-native foundations and alternatives.
Each stance contains finer subcodes. We code the claimed primary contribution,
not every activity reported in the paper. For genuine ties between
\textsc{Interrogate} and another stance, we assign \textsc{Interrogate}. This
gives the rare category the benefit of the doubt and makes its observed scarcity
a conservative result. Stance identifies the position of a contribution; it
does not measure its technical difficulty, intellectual depth, or quality.

\emph{Orientation.}
Finally, we distinguish \textsc{AI4SE}, in which AI is used to address an SE
problem, from \textsc{SE4AI} (other direction). Orientation is orthogonal to stance: a paper can,
for example, service either an AI4SE technique or an SE4AI system.

\subsection{Coding procedure and reliability}
\label{sec:coding-reliability}

We developed the codebook iteratively and froze its definitions and boundary
rules before producing the final population labels. The procedure follows the
SIGSOFT empirical standards' emphasis on explicit constructs, traceable
decisions, and reproducible analysis~\cite{ralph2021empirical}. A model coder
then applied the frozen codebook to the corpus in bounded batches. For every
paper, the coding record retained the assigned labels, supporting title--abstract
evidence, a decision rationale, and a confidence judgment; borderline cases
were explicitly flagged. We used title and abstract evidence consistently
across the corpus. Claims requiring implementation, full-text, or citation-level
evidence remain outside the resolution of the study.

Prior work shows that language models can perform some structured annotation
tasks effectively~\cite{gilardi2023chatgpt}, but this does not make their labels
ground truth. Recent FSE work similarly treats AI as a research assistant whose
outputs remain subject to human assessment: Liang et al.\ evaluated
GPT-4-generated research analyses with software-engineering experts
\cite{liang2024replicate}, while AISysRev combines model-based title--abstract
screening with human review of difficult boundary cases
\cite{huotala2026aisysrev}. We therefore assess the model coder against
independent human judgments and report agreement rather than assuming
correctness.

Two human coders independently coded the same 100 screened candidates for the
SE problem and the three-way research posture, without access to the population
labels or to each other's decisions. Eight candidates were subsequently
classified as non-AI, and four additional records lacked a complete label from
one coder, leaving 88 AI papers with complete paired judgments.
Human--human agreement was $65.9\%$ for the 22-way problem vocabulary
(Cohen's $\kappa=.624$) and $86.4\%$ for research posture
($\kappa=.680$). On the same 88 papers, agreement between the model and the two
human coders ranged from $67.0\%$ to $81.8\%$ for the problem code
($\kappa=.640$--$.801$) and from $83.0\%$ to $86.4\%$ for posture
($\kappa=.588$--$.679$). We report these pairwise comparisons separately rather
than constructing a post-hoc consensus label. Subcodes, lineage, and orientation
were not included in this double-coding exercise; analyses using those
dimensions are therefore interpreted with corresponding caution.
We further evaluated the model annotations against a random sample of 350 human-coded cases.
Human and model labels achieved substantial agreement
(Cohen's $\kappa = 0.72$), supporting their use for population-level
distributional analysis.

\subsection{Corpus analysis}

We analyze the corpus descriptively because it is the full population of the
selected research tracks rather than a probability sample. We calculate annual
paper counts, prevalence relative to all eligible papers, technology-family
composition, and average annual output before and during the LLM wave. Capacity
comparisons use the complete periods 2018--2022 and 2023--2025; 2026 is retained
only in analyses for which partial coverage is explicitly marked.

We compare changes across venues, the eight macro areas, and the 22
fine-grained problem codes. To separate redistribution from the overall growth
of AI research, we also express each problem as a share of the AI papers within
its wave and report percentage-point changes between waves. We summarize
lineage overall and by year, cross-tabulate lineage with orientation, and
compare stance both within the AI corpus and as a proportion of all
research-track output. All denominators are stated with the corresponding
result. These comparisons identify expansion, contraction, and compositional
redistribution in published output; without submission and acceptance data,
they do not establish that AI research causally displaced non-AI work.

\subsection{Community survey}

We use the publicly released ICSE 2026 FoSE pre-survey, which contains 280
completed responses from members of the SE research community
\cite{fose2026survey}. We selected three open-ended questions that form a
diagnosis--future--response sequence: what currently does not work in the
community (Q8), what respondents fear for the next generation (Q12), and the
one change they would make (Q9). None mentions AI, LLMs, or automation. AI
content in the answers is therefore respondent-introduced rather than
prompted.

After retaining the released non-empty and non-substantive answers so that
denominators remain traceable to the source, the dataset contains 656 answers
from 244 respondents: 233 for Q8, 208 for Q12, and 215 for Q9. The unit of
analysis is one respondent's answer to one question.

Each answer received a question-specific primary thematic code and, when a
distinct second idea was present, an optional secondary code. Primary codes
capture the dominant or most elaborated theme. Independently of that thematic
code, we classified the answer's relationship to AI as \textsc{Absent},
\textsc{Concern}, \textsc{Opportunity}, \textsc{Mixed}, or \textsc{Neutral}.
This cross-cutting label was assigned from the response text itself, not inferred
from keywords or from the answer's primary theme. We retained a supporting
evidence span, rationale, and confidence judgment for each decision. The
analysis reports theme frequencies by question, AI salience and valence by
question, and respondent-level counts after deduplicating respondents across the
three questions.

The survey coding was produced by a single model coder. As an internal stability
check, the same model recoded the primary theme of a stratified, label-withheld
sample of 150 answers and reproduced 147 labels ($98.0\%$;
$\kappa=.979$). Because the second pass used the same model after prior exposure
to the corpus, this is an optimistic estimate of self-consistency, not
independent inter-rater reliability or evidence of accuracy. We therefore use
the survey to provide qualitative resonance and illustrative respondent
accounts, not to estimate the prevalence of these views in the wider SE
community.


\section{Results}
\label{sec:results}

We present the results in four stages: how AI changed the scale and distribution
of SE research (RQ1), how the LLM wave changed the problems pursued (RQ2), how the field
engages with AI (RQ3), and how researchers perceive and respond to this shift (RQ4).

\subsection{How has AI changed the scale and distribution of SE research? \textit{(RQ1)}}
\label{sec:rq1}

The growing prevalence of AI papers establishes their visibility, but not how
they have changed the field's publication portfolio. A rising AI share could
reflect an expansion in total research output, a contraction of non-AI work, or
an uneven redistribution across venues and problem areas. We distinguish among
these possibilities by tracing the complete research-track output of ICSE, FSE,
and ASE from 2018 to 2026.

\begin{figure*}[t]
  \centering
  \begin{subfigure}[b]{0.40\textwidth}
    \centering
    \includegraphics[width=\linewidth]{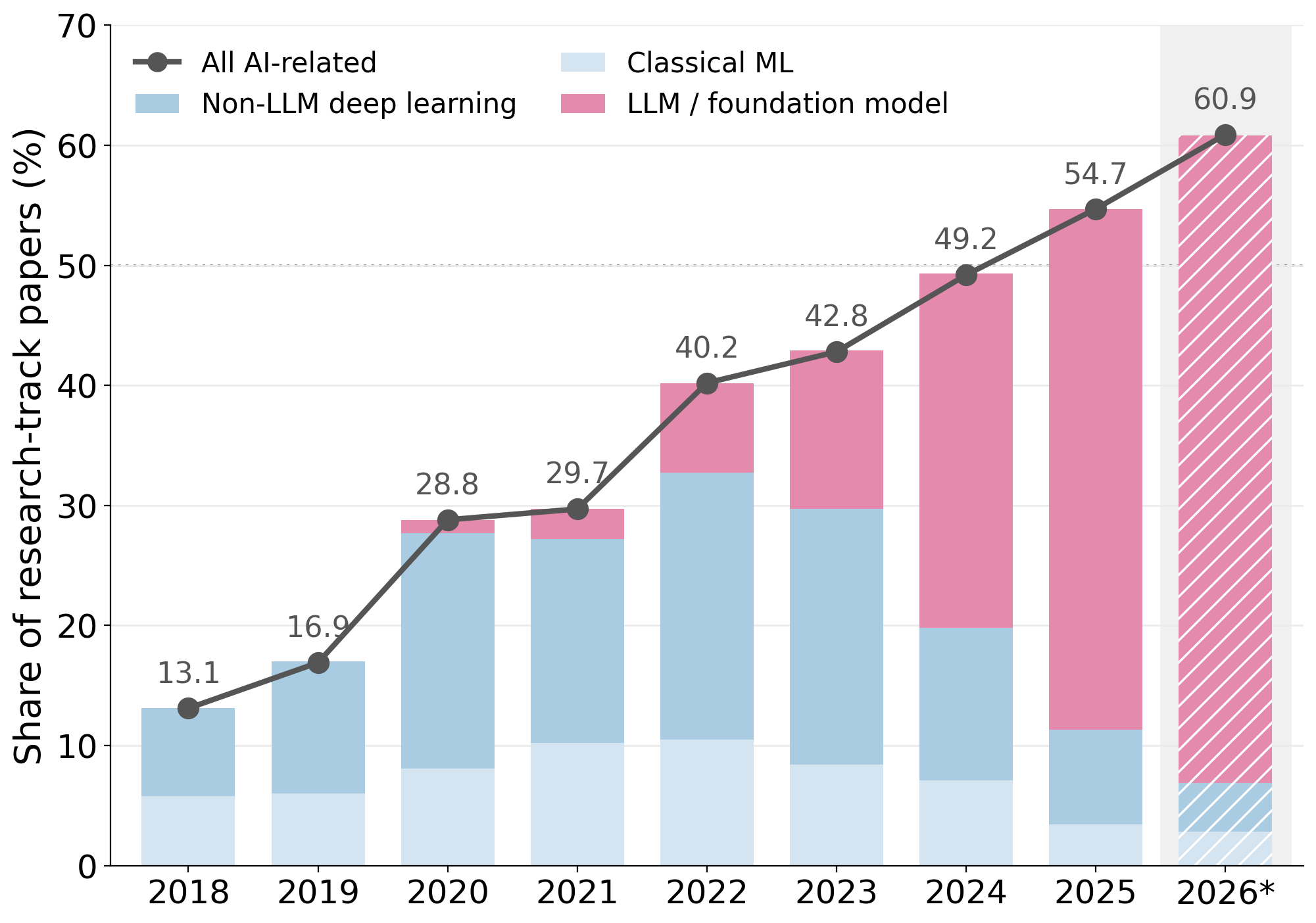}
    \caption{Prevalence \& composition}
    \label{fig:rq1-prevalence}
  \end{subfigure}\hfill
  \begin{subfigure}[b]{0.20\textwidth}
    \centering
    \includegraphics[width=\linewidth]{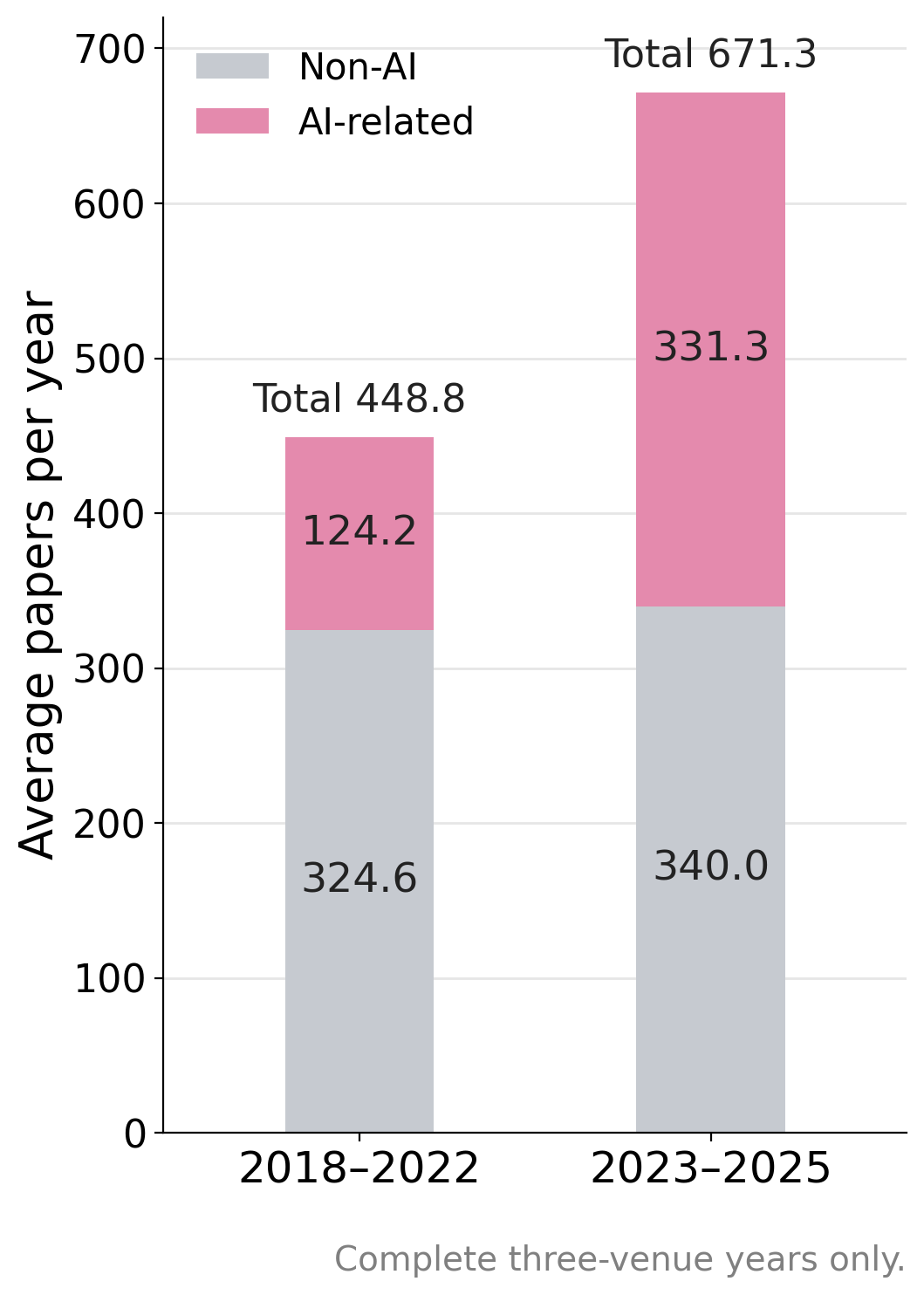}
    \caption{Publ. capacity}
    \label{fig:rq1-capacity-growth}
  \end{subfigure}\hfill
  \begin{subfigure}[b]{0.36\textwidth}
    \centering
    \includegraphics[width=\linewidth]{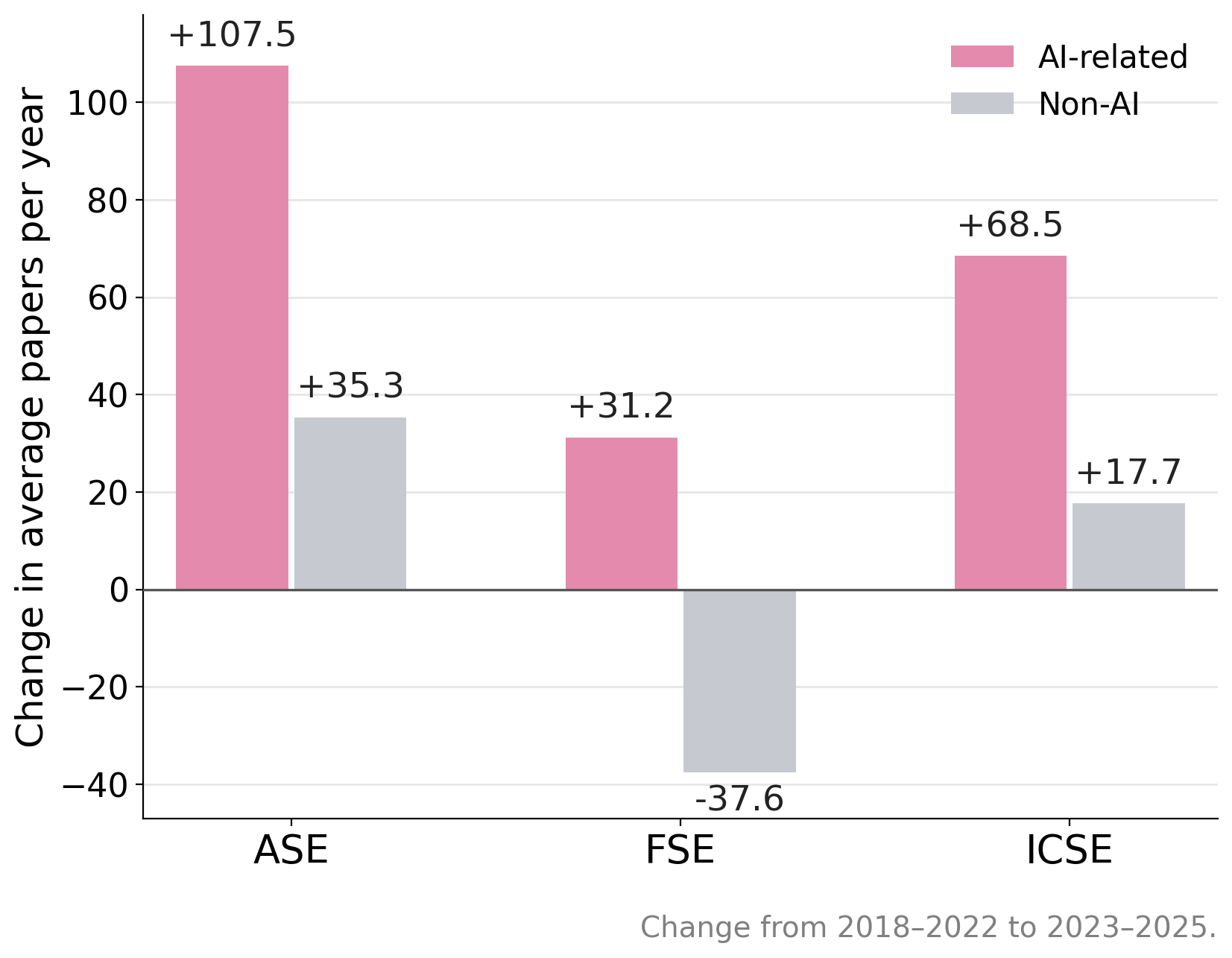}
    \caption{Venue-level change}
    \label{fig:rq1-venues}
  \end{subfigure}
  \caption{How AI changed the scale and distribution of SE research in the ICSE,
  FSE, and ASE research tracks. \textbf{(a)} AI prevalence and technological
  composition, 2018--2026; technology families are mutually exclusive (LLM $>$
  non-LLM deep learning $>$ classical ML), the line marks the overall AI-related
  share, and the $^{*}$2026 bar is partial because ASE is unavailable.
  \textbf{(b)} Average annual research-track output before (2018--2022) and
  during (2023--2025) the LLM wave; AI accounts for $93.1\%$ of the increase in
  annual volume. \textbf{(c)} Change in average annual AI and non-AI output by
  venue, 2018--2022 to 2023--2025; ASE and ICSE grew both categories, while
  FSE's AI gain coincided with a larger non-AI loss.}
  \Description{Three panels in one row: a stacked bar chart of AI share rising to
  54.7\% by 2025, a two-bar comparison of average annual output before and during
  the LLM wave, and a grouped bar chart of AI and non-AI change per venue.}
  \label{fig:rq1-overview}
\end{figure*}

If this concern is reflected in the publication record, its most direct
signature should be compositional: AI should claim a growing share of research
output, rather than merely produce more papers as the conferences themselves
expand. We therefore begin with prevalence---the proportion of all
research-track papers that study or use AI. We further separate AI papers by
technology family to determine whether any increase represents the gradual
accumulation of AI research or the arrival of a distinct technological wave.

Figure~\ref{fig:rq1-prevalence} shows a transformation that is difficult to
attribute to conference growth alone. AI-related work rose from 50 of 382
research-track papers ($13.1\%$) in 2018 to 421 of 769 ($54.7\%$) in 2025. By
then, more than half of the research published at ICSE, FSE, and ASE was
AI-related. Yet the aggregate rise conceals a change in its technological basis.
Non-LLM deep learning expanded through 2022, but subsequently receded as large
language and foundation models rose in its place. In 2025, LLM papers alone
constituted $43.4\%$ of all research-track papers and $79.3\%$ of AI papers.
What appears in the aggregate as the continued growth of AI is therefore better
understood as two successive waves, with the LLM wave rapidly overtaking the
deep-learning research that preceded it. The partial 2026 corpus continues this
trajectory, reaching $60.9\%$ AI; however, because ASE 2026 is not yet
represented, we exclude that year from comparisons with complete conference
years.


Prevalence alone, however, cannot tell us whether AI grew by enlarging the
publication portfolio or by replacing non-AI research within a fixed capacity.
If the growth was primarily additive, AI output should rise without a
corresponding decline in non-AI work. If it instead reflected a reallocation of
existing publication capacity, the rise in AI should coincide with a contraction
in non-AI output. To distinguish these possibilities, we compared average annual
output before the LLM wave (2018--2022) with the complete years of the LLM wave
(2023--2025).

Research-track output increased from 448.8 to 671.3 papers per year
(Figure~\ref{fig:rq1-capacity-growth}). Of the additional 222.5 papers published
annually, 207.1 were AI-related, while non-AI output increased by only 15.4.
AI therefore accounted for $93.1\%$ of the observed increase in publication
volume.

At the level of the three-venue system, this pattern points primarily to
expansion rather than replacement: the venues accommodated a substantial
increase in AI research without an aggregate contraction in non-AI output.
Yet aggregate stability does not imply that every part of the system remained
stable. Growth in one venue could conceal contraction in another. We therefore
next ask whether ICSE, FSE, and ASE accommodated the rise of AI in the same way.

The venue-level comparison reveals two distinct patterns
(Figure~\ref{fig:rq1-venues}). At ASE, average annual output increased by 107.5
AI-related papers and 35.3 non-AI papers; at ICSE, it increased by 68.5 and 17.7,
respectively. In both venues, the expansion of AI research occurred alongside
continued growth in non-AI work. FSE followed a different trajectory. Its
average annual AI output increased by 31.2 papers, while its non-AI output
decreased by 37.6. The system-wide pattern of expansion, therefore, conceals a
venue-level contraction: at FSE, the decline in non-AI papers was greater than
the increase in AI papers.

These divergent trajectories preclude a simple characterization of the
shift. The evidence does not indicate field-wide displacement, since ASE and
ICSE expanded both its AI and non-AI portfolios. Nor does it indicate that AI
was uniformly layered onto an otherwise unchanged publication base, since FSE's
growth in AI coincided with a substantial contraction in non-AI work. Instead,
expansion and compositional reallocation occurred simultaneously in different
parts of the venue system. This conclusion is limited to the composition of
accepted papers. Without corresponding submission and acceptance data, we
cannot determine whether the observed reallocation resulted from AI research
causally crowding out non-AI work.


The venue-level analysis identifies where publication growth occurred
institutionally, but it does not show how that growth altered the distribution of
attention across software engineering problems. In particular, the near-stability
of aggregate non-AI output could conceal a substantial topical redistribution:
growth in some problem areas may have offset contraction in others. To examine
this possibility, we grouped the corpus into eight broad SE problem areas and
compared their average annual output before and during the LLM wave.

\begin{figure*}[t]
  \centering
  \begin{subfigure}[b]{0.52\textwidth}
    \centering
    \includegraphics[width=\linewidth]{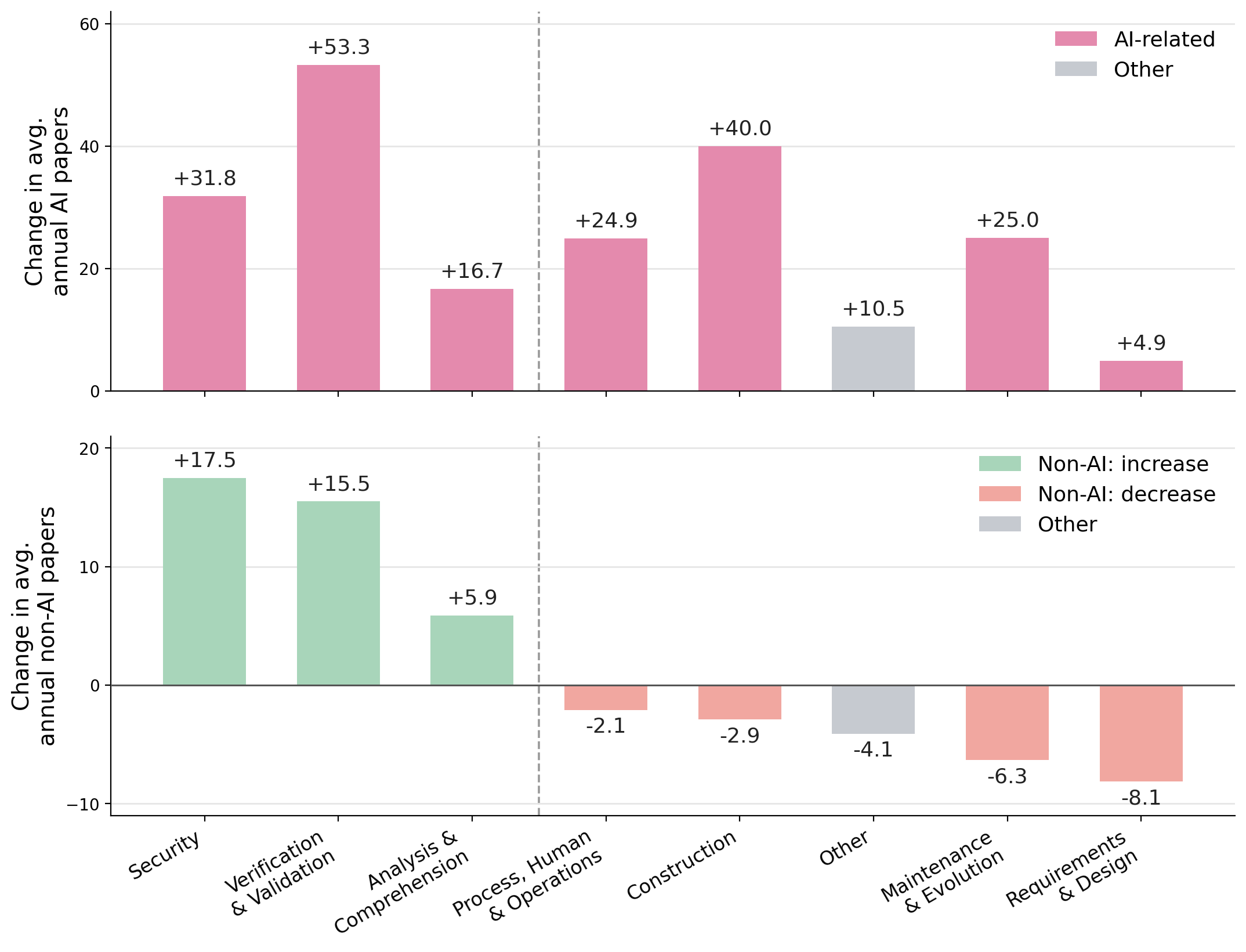}
    \caption{Macro problem areas}
    \label{fig:rq1-macro-areas}
  \end{subfigure}\hfill
  \begin{subfigure}[b]{0.455\textwidth}
    \centering
    \includegraphics[width=\linewidth]{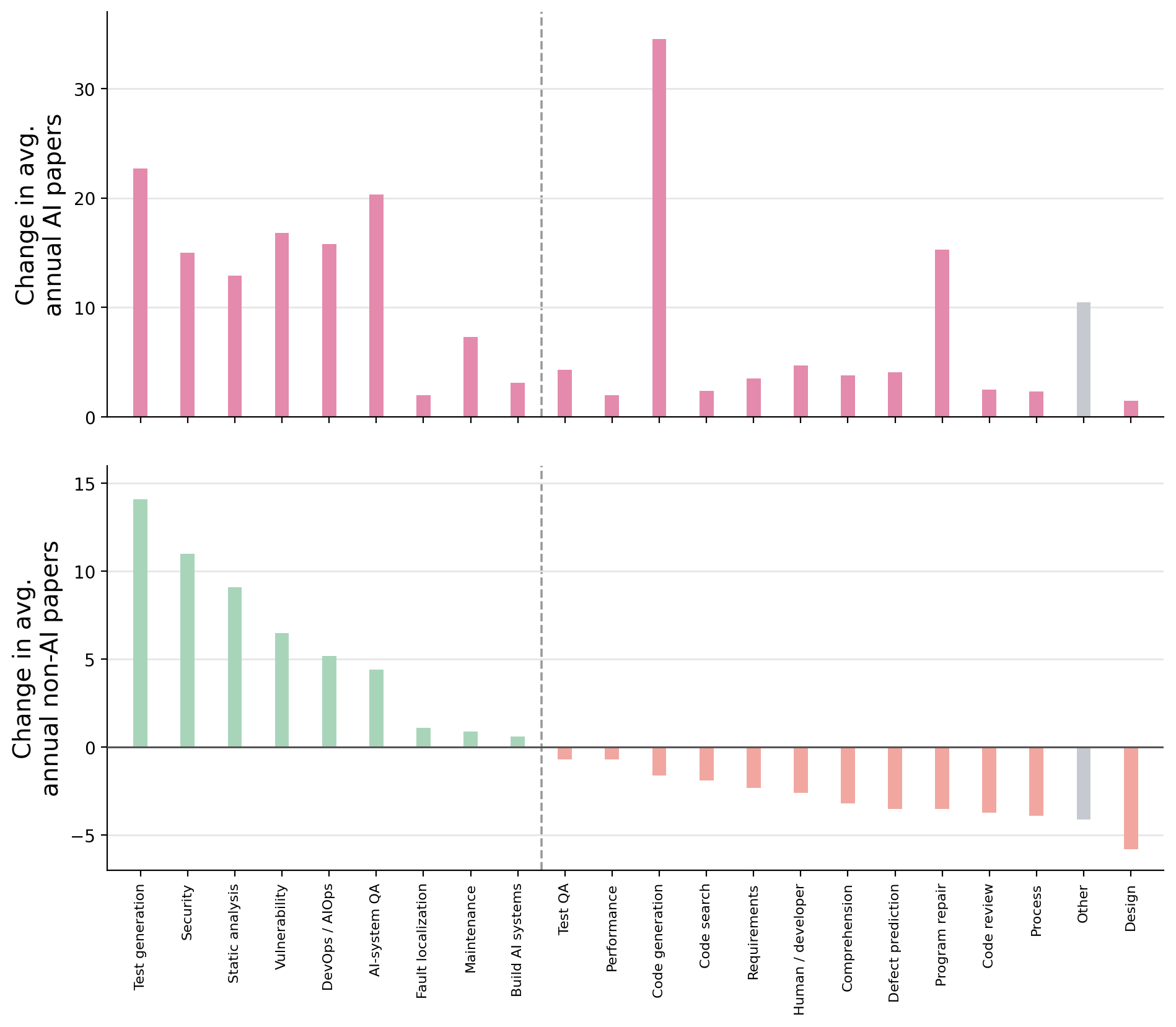}
    \caption{Fine-grained problem areas}
    \label{fig:rq1-problems}
  \end{subfigure}
  \caption{Change in average annual AI and non-AI research-track output,
  2018--2022 to 2023--2025. \textbf{(a)} Across the eight macro problem areas;
  the dashed divider separates areas in which non-AI output expanded from those
  in which it contracted, and the residual \textsc{Other} category is shown in
  gray. \textbf{(b)} Fine-grained decomposition across 22 problem codes; AI
  output increased in all 21 substantive areas, while non-AI output increased in
  nine and decreased in twelve.}
  \Description{Two panels. Left: paired bars of AI and non-AI change across eight
  macro areas, AI positive everywhere, non-AI rising in three areas and falling
  in the rest. Right: the same comparison across 22 fine-grained problem codes,
  with AI positive in all substantive areas and non-AI mixed in sign.}
  \label{fig:rq1-macro}
\end{figure*}

The resulting pattern is asymmetric (Figure~\ref{fig:rq1-macro}). AI output
increased across every macro area, indicating that its growth was not confined
to a small set of applications. Non-AI output, however, increased in only three
of the seven substantive areas: Security, Verification and Validation, and
Analysis and Comprehension. It declined in Process, Human and Operations;
Construction; Maintenance and Evolution; and Requirements and Design. The
contrast is clearest between Verification and Validation, which gained 53.3 AI
and 15.5 non-AI papers per year, and Requirements and Design, which gained only
4.9 AI papers while losing 8.1 non-AI papers per year.

The aggregate stability of non-AI research therefore does not represent a stable
underlying portfolio. It is the net result of expansion in three broad areas
offsetting contraction in four others. AI research spread across the entire
agenda, while the remaining non-AI portfolio became internally redistributed.
We report the residual \textsc{Other} category for completeness but do not
interpret it as a coherent research area.

Broad areas can still conceal opposing changes among the problems they contain.
We therefore decomposed the same comparison into 22 fine-grained problem codes
(Figure~\ref{fig:rq1-problems}). Excluding \textsc{Other}, AI output increased
in all 21 substantive problem areas, whereas non-AI output increased in nine and
decreased in twelve. The largest increases in AI output occurred in code
generation ($+34.5$ papers per year), test generation ($+22.7$), and AI-system
QA ($+20.3$). Yet even these rapidly growing areas followed different non-AI
trajectories. Test generation gained 14.1 non-AI papers per year, whereas code
generation lost 1.6. Similarly, program repair gained 15.3 AI papers while
losing 3.5 non-AI papers per year. Design experienced the largest contraction
in non-AI output ($-5.8$), despite gaining only 1.5 AI papers per year.

These contrasts show why neither prevalence nor aggregate publication capacity
is sufficient to characterize the change. AI expanded throughout the problem
portfolio, but that expansion was associated with different changes in the
surrounding literature: some problems attracted more AI and non-AI work
simultaneously, while others gained AI papers as their non-AI output contracted.
The result is not a wholesale collapse of non-AI research, but a selective
redistribution of publication attention concealed by aggregate growth.


Taken together, AI did not merely become more prevalent in SE research. It became the
principal source of the field's growth, accounting for $93.1\%$ of the observed
increase in publication volume and expanding across every substantive problem
area. Yet this expansion was not neutral across the existing portfolio. Growth
at ASE, ICSE, and several problem areas offset contraction at FSE and in twelve
of the twenty-one fine-grained problems. The field grew through AI, but the
portfolio beneath that growth was being redistributed.


\subsection{How did the LLM wave reshape what SE research studies? \textit{(RQ2)}}
\label{sec:rq2}

\begin{table}[b]
  \centering
  \caption{Problem lineage of LLM-wave AI papers by year: paper count with its
  share of that year's AI papers in brackets. Established problems dominate
  throughout; the \textsc{New} sliver grows but stays small.}
  \label{tab:rq2-lineage}
  \begin{tabular}{l
      >{\columncolor{lineEst}}c
      >{\columncolor{lineRef}}c
      >{\columncolor{lineNew}}c
      r}
    \toprule
    Year & \textsc{Established} & \textsc{Reframed} & \textsc{New} & $n$ \\
    \midrule
    2023 & $259$ ($97.4\%$) & $2$ ($0.8\%$) & $5$ ($1.9\%$) & $266$ \\
    2024 & $285$ ($92.8\%$) & $0$ ($0.0\%$) & $22$ ($7.2\%$) & $307$ \\
    2025 & $381$ ($90.5\%$) & $4$ ($1.0\%$) & $36$ ($8.6\%$) & $421$ \\
    2026 & $296$ ($91.4\%$) & $1$ ($0.3\%$) & $27$ ($8.3\%$) & $324$ \\
    \midrule
    \textbf{Pooled} & $\mathbf{1{,}221}$ ($\mathbf{92.6\%}$)
                    & $\mathbf{7}$ ($\mathbf{0.5\%}$)
                    & $\mathbf{90}$ ($\mathbf{6.8\%}$)
                    & $\mathbf{1{,}318}$ \\
    \bottomrule
  \end{tabular}
\end{table}

RQ1 established the scale and location of the shift, but not its intellectual
direction. In particular, prevalence cannot tell us whether the LLM wave opened
new software engineering problems, reframed existing ones, or applied a new
technological capability to an inherited research agenda. To distinguish among
these possibilities, we compared the SE problem posed by each LLM-wave paper
against the problem distribution observed during the deep-learning wave. We
classified each problem's lineage as \textsc{Established}, \textsc{Reframed},
or \textsc{New} (\S\ref{sec:method}).

The lineage distribution is overwhelmingly \textsc{Established}
(Table~\ref{tab:rq2-lineage}). Of the $1{,}318$ LLM-wave papers, $92.6\%$
address a problem already present in the deep-learning wave, while only $6.8\%$
introduce a \textsc{New} problem and $0.5\%$ \textsc{Reframe} an existing one.
This pattern persists throughout the LLM wave. The share classified as
\textsc{New} rises from $2\%$ in 2023 to approximately $8$--$9\%$ in 2025 and
the partial 2026 corpus, but never exceeds one in ten papers. The LLM wave
therefore introduced relatively little new problem territory; most papers
applied a new technological family to problems the field had already established.

This result describes the boundaries of the problem portfolio, but not the
distribution of attention within it. Even if the LLM wave inherited nearly all
of its problems, it may still have elevated some of them while others receded.
We therefore next examine how the relative composition of the AI problem
portfolio changed between the deep-learning and LLM waves.


\begin{figure*}[t]
  \centering
  \begin{subfigure}[t]{0.52\textwidth}
    \centering
    \includegraphics[width=\linewidth]{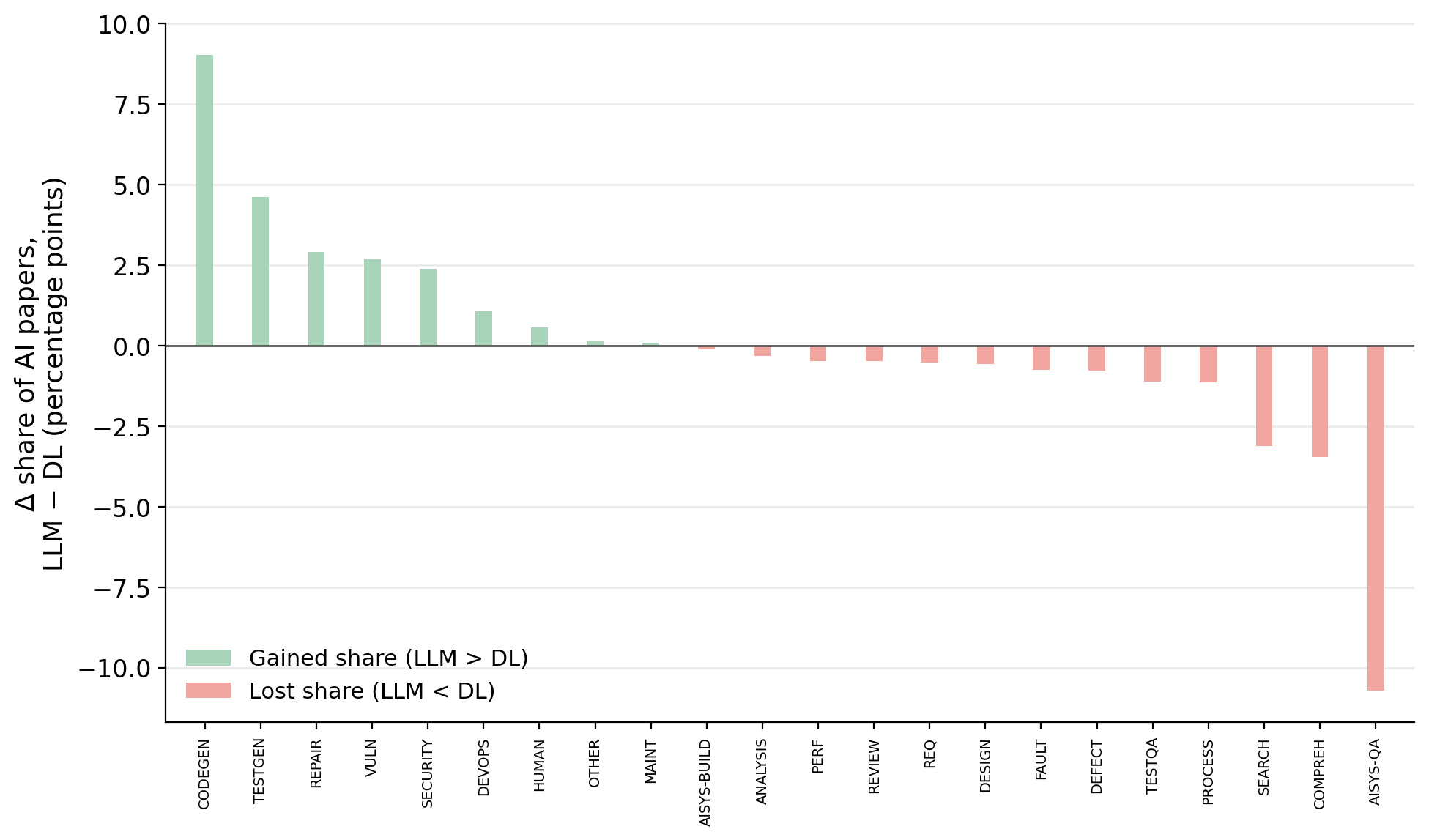}
    \caption{Portfolio shift}
    \label{fig:rq2-portfolio-shift}
  \end{subfigure}\hfill
  \begin{subfigure}[t]{0.46\textwidth}
    \centering
    \includegraphics[width=\linewidth]{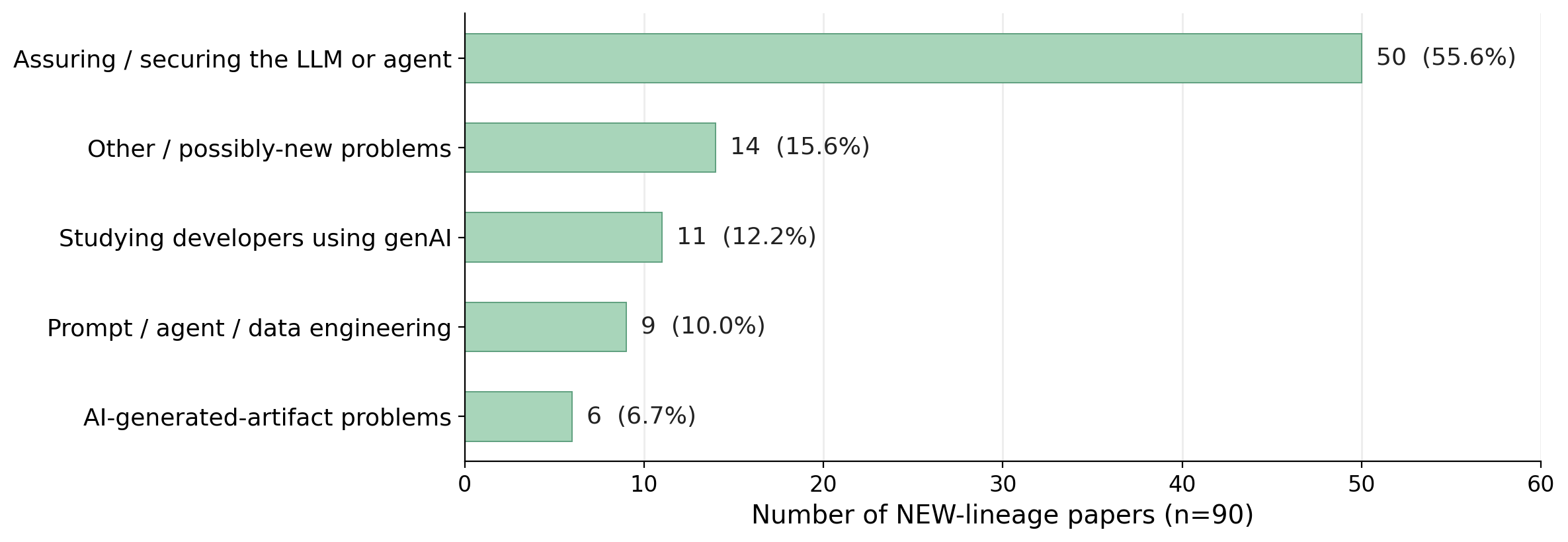}
    \caption{New problems}
    \label{fig:rq2-new}
  \end{subfigure}
  \caption{How the LLM wave reorganized an inherited problem map.
  \textbf{(a)} Change in each problem's share of AI papers from the
  deep-learning wave to the LLM wave (\% points), at the 22-category
  Leg~A granularity; green bars mark problems that gained share under the LLM
  wave ($\text{LLM} > \text{DL}$) and coral bars those that lost share
  ($\text{LLM} < \text{DL}$). \textbf{(b)} The $90$ \textsc{New}-lineage papers
  grouped by the kind of ``new'' problem they pose; nearly all concern the
  model, its outputs, or its use, and at most one poses a software-engineering
  problem independent of the model.}
  \Description{Left: a diverging bar chart of per-problem change in AI-paper
  share. Right: a horizontal bar chart of the 90 new-lineage papers across five
  categories.}
  \label{fig:rq2-portfolio}
\end{figure*}

To capture this change, we treated the AI papers in each wave as a portfolio
distributed across the 22 SE problem categories. For each category, we
calculated the proportion of all AI papers in the deep-learning wave and the
corresponding proportion in the LLM wave
(Figure~\ref{fig:rq2-portfolio}. We use proportions because the LLM wave
contains many more AI papers overall. This comparison shows whether a problem
became more prominent within AI research, rather than merely gaining papers
because the entire AI literature grew.

The comparison reveals a substantial redistribution of attention. Code
generation experienced the largest gain, rising from $3.9\%$ of AI papers in
the deep-learning wave to $12.9\%$ in the LLM wave, an increase of 9.0
percentage points. Test generation ($+4.6$ points), program repair ($+2.9$),
vulnerability detection ($+2.7$), and security ($+2.4$) also became more
prominent. In contrast, quality assurance for AI systems experienced the
largest decline, falling from $25.3\%$ to $14.6\%$ of the AI portfolio.
Comprehension ($-3.5$ points) and code search ($-3.1$) also receded.

These declines indicate a loss of relative prominence, not necessarily a
reduction in the number of papers. \textsc{AI-system QA}, for example, grew from 157 to
192 papers, but it did not grow as quickly as the AI literature surrounding
other problems. The LLM wave, therefore, altered the center of gravity within an
inherited problem map: attention moved toward code generation and other
generative tasks, while the deep-learning wave's emphasis on evaluating AI
systems diminished.

\begin{table}[h!]
  \caption{Problem lineage by orientation, LLM-wave AI papers ($n{=}1{,}318$).
  Counts, with the orientation split within each lineage in parentheses.}
  \label{tab:rq2-lineage-orient}
  \begin{tabular}{lrrr}
    \toprule
    Lineage & \textsc{AI4SE} & \textsc{SE4AI} & Total \\
    \midrule
    \textsc{Established} & $995$ ($81.5\%$) & $226$ ($18.5\%$) & $1{,}221$ \\
    \textsc{Reframed}    & $6$   ($85.7\%$) & $1$   ($14.3\%$) & $7$ \\
    \textsc{New}         & $40$  ($44.4\%$) & $50$  ($55.6\%$) & $90$ \\
    \midrule
    Total                & $1{,}041$ & $277$ & $1{,}318$ \\
    \bottomrule
  \end{tabular}
\end{table}

What, then, were the new problems introduced during the LLM wave? We examined
all 90 papers classified as \textsc{New} and grouped them according to the kind
of problem they posed (Figure~\ref{fig:rq2-new}). Most of the new
problems arose from the demands of building, using, or governing LLMs themselves
rather than software engineering problems independent of the new technology. The
largest group, comprising 50 of the 90 papers ($55.6\%$), addressed the
assurance or security of LLMs and agents, including hallucinations, jailbreaks,
prompt injection, data poisoning, agent failures, fairness, robustness, and
trust. The remaining papers examined developers' use of generative AI ($11$),
the engineering of prompts, agents, and training data ($9$), and problems
associated with AI-generated artifacts ($6$); a residual set of $14$ covered
other, possibly new problems that we could not cleanly attribute to these
categories. Across all five groups, at most one paper posed a new SE problem
that was not predicated on the existence of the model.

The orientation of these papers reinforces this distinction
(Table~\ref{tab:rq2-lineage-orient}). Whereas $81.5\%$ of papers addressing
\textsc{Established} problems are \textsc{AI4SE}, a majority of the
\textsc{New} papers are \textsc{SE4AI} ($55.6\%$, 50 of 90). A paper addressing
a new problem is therefore more than three times as likely to concern the
engineering of an AI system as a paper addressing an established problem
($55.6\%$ versus $18.5\%$). The limited expansion of the problem portfolio was
thus directed primarily toward the new burdens created by LLM-based systems,
rather than toward previously unexamined problems in SE.

To summarize, the LLM wave changed the emphasis of the SE problem portfolio far more than its
boundaries. Most papers returned to problems established during the
deep-learning wave, while attention within that inherited map shifted toward
code generation and other generative tasks. The small amount of new territory
arose primarily from the work required to build, assure, and govern LLM-based
systems themselves. The technological wave reorganized an existing
research agenda around the capabilities and consequences of a new technical
family, rather than replacing it with a new set of software engineering
problems.


\subsection{How does SE research engage with AI? \textit{(RQ3)}}
\label{sec:rq3}

Although RQ2 establishes that most LLM-wave papers revisit inherited problems,
recurrence does not reveal how researchers engage with AI within them. A paper
may apply AI to an SE task, produce knowledge or infrastructure supporting its
use, or interrogate its assumptions and develop SE-native alternatives. We
distinguish these research stances as \textsc{Apply}, \textsc{Service}, and
\textsc{Interrogate}.

\begin{table}[b]
  \centering
  \caption{Research stance by wave. Stance columns report percentages of AI
  papers within each wave; A$+$S/all reports Apply plus Service as a percentage
  of all research-track papers in that wave.}
  \label{tab:rq3-stance}
  \small
  \begin{tabular}{lrrrrr}
    \toprule
    Wave & $n_{\mathrm{AI}}$ & Apply & Service & Inter. & A$+$S/all \\
    \midrule
    DL ($\le$2022)  & $621$    & $82.1$ & $14.3$ & $3.5$ & $26.7\%$ \\
    LLM ($\ge$2023) & $1{,}318$ & $72.4$ & $22.9$ & $4.7$ & $49.3\%$ \\
    \midrule
    Pooled (2018--2026) & $1{,}939$ & $75.5$ & $20.2$ & $4.3$ & $38.4\%$ \\
    \bottomrule
  \end{tabular}
\end{table}

The distribution is highly uneven (Table~\ref{tab:rq3-stance}). Across the
$1{,}939$ AI papers, $75.5\%$ \textsc{Apply} AI and $20.2\%$
\textsc{Service} its use through evaluation, characterization, or supporting
infrastructure. Only $4.3\%$ \textsc{Interrogate} its assumptions, direction,
or disciplinary foundations. Thus, $95.7\%$ of the literature either uses AI
capabilities or supports their use, while fewer than one paper in twenty
questions the paradigm or develops alternatives.



\begin{figure*}[t]
  \centering
  \begin{subfigure}[b]{0.45\textwidth}
    \centering
    \includegraphics[width=\linewidth]{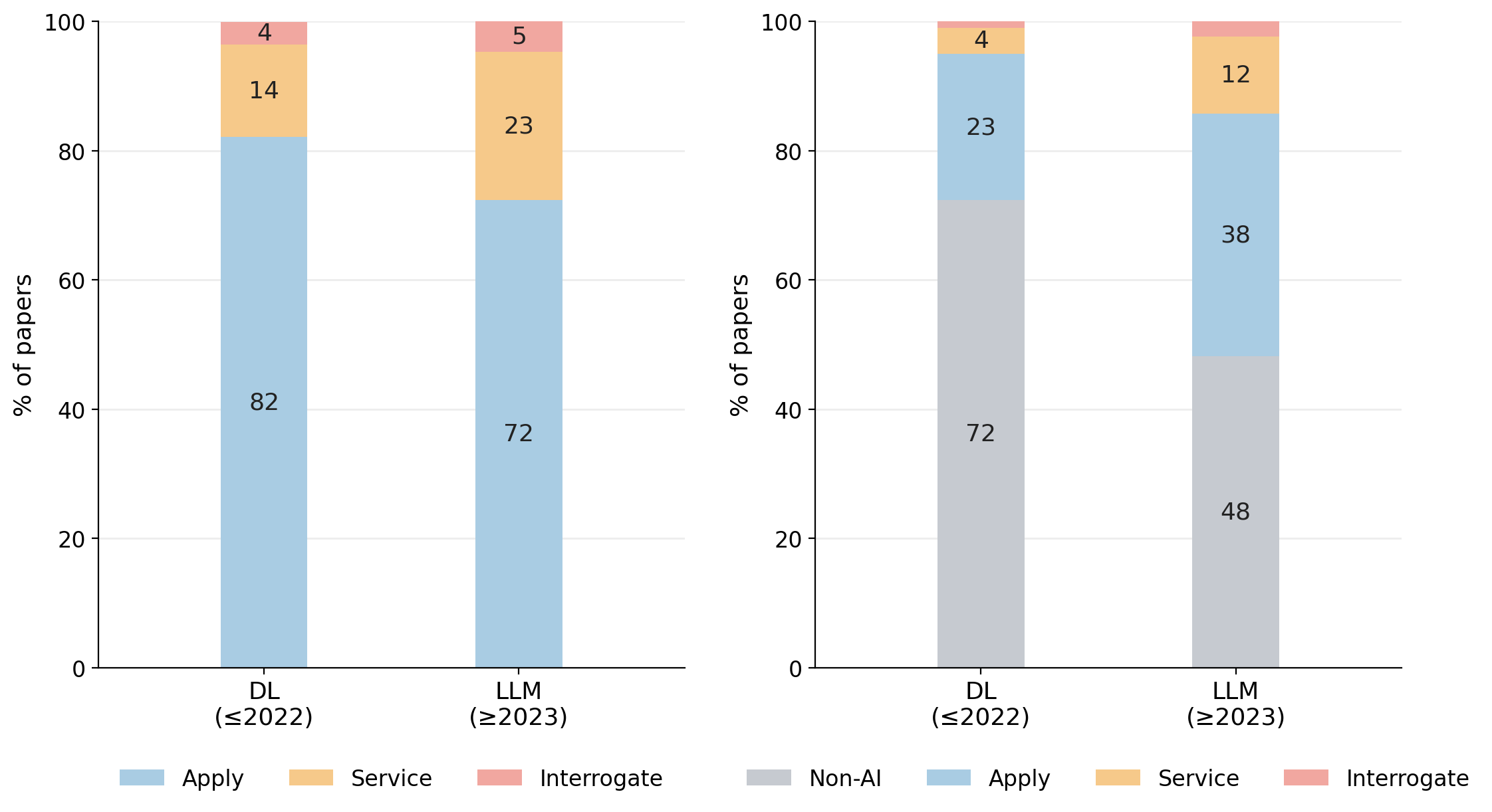}
    \caption{Stance composition by wave}
    \label{fig:rq3-stance-comp}
  \end{subfigure}\hfill
  \begin{subfigure}[b]{0.55\textwidth}
    \centering
    \includegraphics[width=\linewidth]{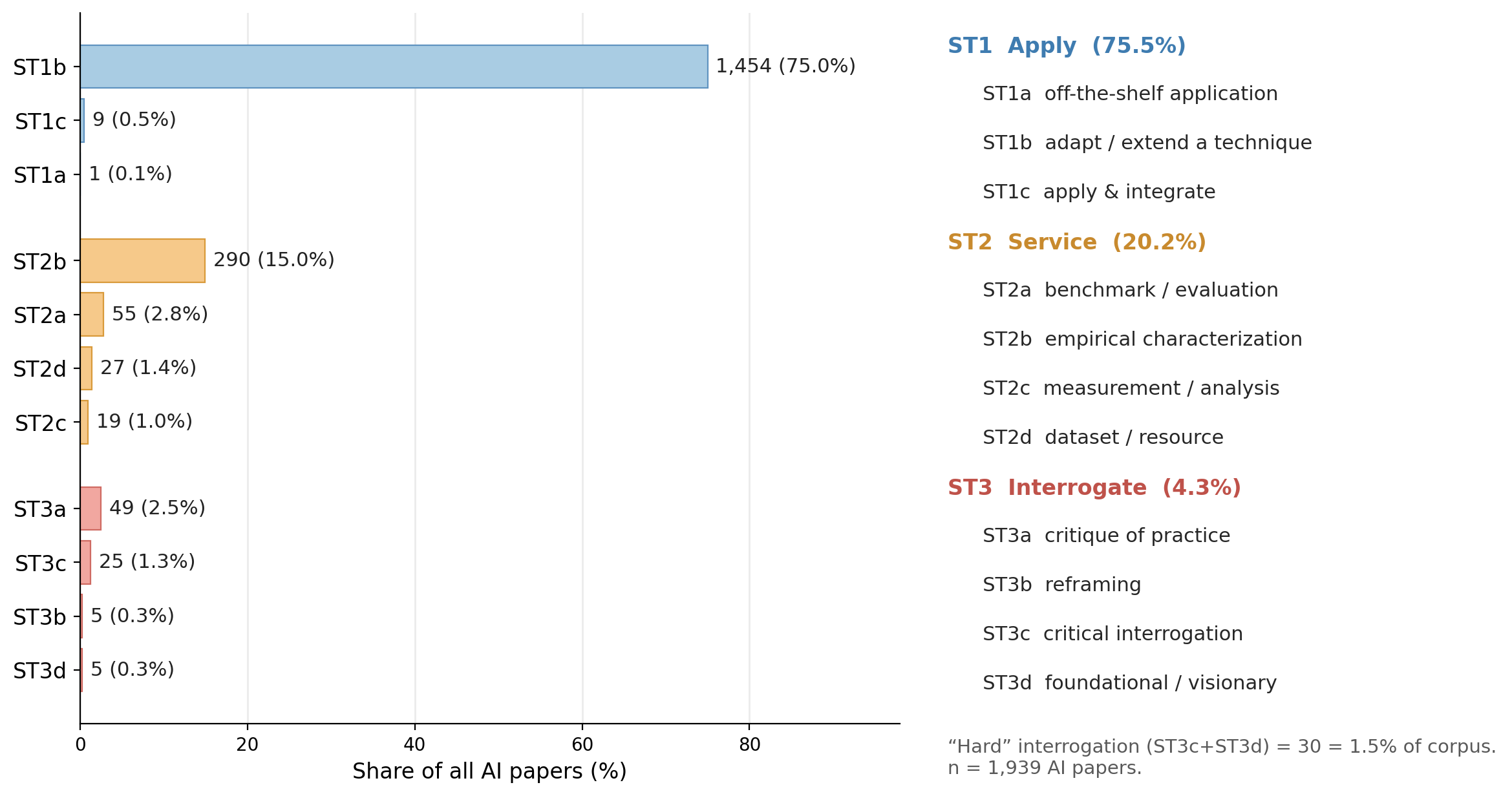}
    \caption{Stance decomposition}
    \label{fig:rq3-stance-decomp}
  \end{subfigure}
  \caption{How SE research engages with AI, across the $1{,}939$ AI papers.
  \textbf{(a)} Research stance by wave. Within AI (left pair), the combined
  \textsc{Apply}--\textsc{Service} block stays $\sim$95--96\% but recomposes from
  application toward service; across all research-track output (right pair),
  apply-and-service work grows from $\sim$27\% to $\sim$49\% while
  \textsc{Interrogate} creeps from $1.0\%$ to $2.4\%$. \textbf{(b)} Decomposition
  of each stance into subcodes: \textsc{Apply} is almost entirely ST1b
  (adapt/extend a technique, $75.0\%$ of the corpus), and the paradigm-level
  interrogation subcodes (ST3c, ST3d) together are just $1.5\%$ of the corpus.}
  \Description{Left: two pairs of 100\% stacked bars comparing the deep-learning
  and LLM waves, one within AI papers and one across all research-track output,
  showing apply and service dominating and interrogation remaining tiny. Right:
  a horizontal bar chart of the eleven stance subcodes with a definition key,
  dominated by ST1b.}
  \label{fig:rq3-stance}
\end{figure*}

The comparison across waves reveals stability at one level and change at
another (Figure~\ref{fig:rq3-stance-comp}). Within AI, the combined
\textsc{Apply}--\textsc{Service} share remains approximately $95$--$96\%$, but
its composition shifts: \textsc{Apply} declines from $82.1\%$ to $72.4\%$,
while \textsc{Service} rises from $14.3\%$ to $22.9\%$.
\textsc{Interrogate} increases only slightly, from $3.5\%$ to $4.7\%$
(Panel~1a). The field therefore shifted attention from constructing AI-based
solutions toward evaluating and supporting them, but not toward questioning
the paradigm. Because AI research itself expanded, \textsc{Apply} and
\textsc{Service} together grew from $26.7\%$ to $49.3\%$ of all
research-track output, while \textsc{Interrogate} grew from only $1.0\%$ to
$2.4\%$ (Panel~1b).

Decomposing the stances shows that they are internally concentrated
(Figure~\ref{fig:rq3-stance-decomp}). ST1b---adapting or extending an AI technique
for an SE task---accounts for $99.3\%$ of \textsc{Apply} and $75.0\%$ of the
entire AI corpus. Application is therefore rarely a bare model call; it more
often constructs a technique around an imported model. \textsc{Service} is
dominated by empirical characterization (ST2b), which represents $74.2\%$ of
service papers, compared with $14.1\%$ for benchmarks or leaderboards and
$6.9\%$ for datasets or research resources. Service work primarily studies how
AI behaves rather than constructing shared evaluation infrastructure.

The small \textsc{Interrogate} stance is weighted toward critique rather than
alternative construction. Of its 84 papers, $58.3\%$ critique current
practices and $29.8\%$ empirically expose limitations of the prevailing
paradigm. Reframing the paradigm and developing SE-native foundations together
account for only ten papers, or $0.5\%$ of the AI corpus. Even when research
steps back from application, it is more likely to identify problems with the
current direction than to formulate an alternative.

Research stance should not be interpreted as intellectual depth or quality.
ST1b includes both thin wrappers and substantial adaptations grounded in
program semantics, execution constraints, or other SE knowledge. Our coding
captures where a paper positions its primary contribution relative to AI;
distinguishing the depth of that contribution requires evidence beyond titles
and abstracts (\S\ref{sec:limitations}).

\subsection{How does the community perceive the AI shift? \textit{(RQ4)}}
\label{sec:rq4}

RQ1--RQ3 traced the AI shift through the published record. To examine how
researchers themselves perceive this transformation, we analyzed three
open-ended questions from the ICSE 2026 FoSE pre-survey~\cite{fose2026survey}:
what currently does not work in the community (Q8), what respondents fear for
the next generation (Q12), and the one change they would make (Q9). None of the
questions mentioned AI; any discussion of it was introduced by respondents.

\emph{What does not work now?}
Respondents' diagnoses centered on the institutions and direction of research.
Peer review and publication was the most frequent primary theme ($61/233$,
$26.2\%$), closely followed by research agenda and relevance ($58/233$,
$24.9\%$). Community and access ($16.7\%$), incentives and metrics ($12.0\%$),
and methodological rigor ($7.3\%$) followed. AI was the primary theme of only 15
answers ($6.4\%$), although it also appeared within complaints about agenda and
rigor. One respondent criticized studies of the form ``we tried something with
an LLM, here it is,'' warning of an ``AI bubble'' within the community
(Respondent 195811076, Q8). AI was therefore not the dominant account of what
currently ails the field; it appeared through longer-standing concerns about
what research is valued and how it is evaluated.

\emph{What threatens the field's future?}
The emphasis changed when respondents considered the next generation.
Publication pressure and wellbeing remained the largest category ($46/208$,
$22.1\%$), but AI-driven change followed closely ($42/208$, $20.2\%$).
Careers and employment accounted for $13.0\%$, while research purpose, skills
and training, and research integrity each accounted for approximately one in
ten answers. These concerns extended beyond employment to expertise and agenda
breadth. One respondent feared that researchers would become
``AI-dependent prompt scribes'' (Respondent 196065978, Q12), while another
warned that rapid AI development leaves ``no time to keep up to date on other
SE advances'' (Respondent 194594580, Q12).

\emph{What should the community change?}
These future concerns did not translate directly into respondents' proposed
reforms. Review and publication reform dominated Q9 ($76/215$, $35.3\%$),
followed by changes to the research agenda ($54/215$, $25.1\%$). Only three
answers ($1.4\%$) made AI-specific reform their primary proposal. One called on
the community to ``increase rigor on papers using LLMs,'' noting that many fail
to compare against non-LLM solutions (Respondent 196098713, Q9).

Because AI could appear without being an answer's primary theme, we also coded
it as a cross-cutting dimension. Table~\ref{tab:rq4-ai-salience} shows whether
each answer raised AI and whether it expressed an AI-related concern.

\begin{table}[!h]
  \centering
  \caption{Unprompted AI salience across the three FoSE survey questions.}
  \label{tab:rq4-ai-salience}
  \small
  \begin{tabular}{llrr}
    \toprule
    Question & Focus & AI raised & AI concern \\
    \midrule
    Q8  & Current diagnosis & $46/233$ ($19.7\%$) & $32/233$ ($13.7\%$) \\
    Q12 & Future concern    & $65/208$ ($31.2\%$) & $44/208$ ($21.2\%$) \\
    Q9  & Desired change    & $15/215$ ($7.0\%$)  & $5/215$ ($2.3\%$) \\
    \bottomrule
  \end{tabular}
\end{table}

Overall, 98 of 244 respondents ($40.2\%$) raised AI at least once, and 69
($28.3\%$) expressed a concern. Its salience, however, was concentrated in the
future: nearly one-third of Q12 answers raised AI, compared with one-fifth of
current diagnoses and only $7.0\%$ of proposed changes.

The concerns themselves were not monolithic. Some questioned the rigor of the
research being produced, describing papers whose contribution amounted to
``LLM/Agent for XYZ'' (Respondent 196053996, Q12) or ``we have thrown an LLM at
a problem, and it mostly worked'' (Respondent 195853386, Q8). Others focused on
the allocation of research attention, warning that ``AI-slop'' was ``crowding
out more meaningful research'' (Respondent 192109802, Q12) and that good work in
other areas was being lost amid the current emphasis on AI (Respondent
194594580, Q8). A third concern involved the integrity of research itself:
respondents anticipated ``fake papers, fake reviews, fake data'' and increasing
reliance on LLMs throughout publication (Respondent 195830557, Q12). Others
worried about the development of expertise. They feared that dependence on
ChatGPT could weaken ``creativity and critical-thinking skills'' (Respondent
195830783, Q12), or leave researchers as ``AI-dependent prompt scribes'' without
command of underlying methods (Respondent 196065978, Q12). Finally, some raised
the material costs of the shift, calling for greater attention to the energy
and other resources consumed by LLM research (Respondent 192867648, Q8).
AI-related concern encompassed research quality, portfolio crowding,
scientific integrity, expertise, disciplinary identity, and environmental cost.

The survey reveals an asymmetry between recognition and response.
Among these respondents, AI is a prominent source of uncertainty about the
field's future, but it has not produced a comparably visible agenda for
collective action. This does not imply indifference: Q9 permitted only one
proposed change, and the FoSE sample is not representative of the entire
community. The narrower result is that AI has entered the community's account of
its future more clearly than its stated agenda for change. This pattern
resonates with RQ3: the field applies and services AI far more often than it
interrogates its premises or develops alternatives.

\section{Discussion}
\label{sec:discussion}

The LLM wave neither displaced non-AI research wholesale nor left the field
unchanged. Instead, it supplied nearly all recent publication growth while
redistributing attention across venues and established problems. Within that
larger AI portfolio, service work grew, but application remained dominant and
interrogation remained rare. The central transformation is therefore not simply
that SE publishes more AI research. A substantial part of the field has been
reorganized around a new technological family without a comparable expansion
in its problem boundaries or in research questioning the direction of that
reorganization.

\subsection{A technological revolution without a problem revolution}

More than nine in ten LLM-wave papers address problems already present during
the deep-learning wave. LLMs nevertheless changed their relative appeal:
generative tasks such as code and test generation gained attention, while other
established areas receded. The transition changed the field's instruments and allocation of effort more than the set of
problems it recognizes. Existing reviews document the breadth of this
redistribution across SE tasks~\cite{fan2023large,hou2024large}; our census
shows how it altered those tasks' position within the wider publication
portfolio.

This pattern is consistent with, but does not establish, capability-driven
agenda formation. Accessible models, benchmarks, and visible performance gains
may make some problems appear more feasible or publishable. Equally,
researchers may begin with important SE problems and select LLMs because they
are appropriate tools. Accepted-paper metadata cannot distinguish these
pathways. The portfolio-level question nevertheless remains important: even
when individual papers are strong, repeated investment in problems aligned
with a dominant technology can narrow collective attention. A similar tension
has been observed beyond SE, where AI use has been associated with increased
individual productivity but reduced collective topical breadth
\cite{hao2026artificial}. Research agendas should therefore be evaluated not
only through individual contributions, but also through the problems that the
resulting portfolio repeatedly rewards or leaves peripheral.

\subsection{Problem continuity is not knowledge continuity}

The recurrence of established problems creates an opportunity for cumulative
research, but does not demonstrate that such accumulation occurred. Two papers
may both study program repair or test generation while sharing few constructs,
baselines, assumptions, or prior findings. A technological transition can
preserve a problem's label while resetting how that problem is formulated and
evaluated. Reviews of LLM-for-SE research already identify fragmented
benchmarks, evaluation practices, and reporting conventions
\cite{hou2024large,wang2023softwaretesting}; our lineage analysis cannot show
whether individual papers recover the strongest knowledge produced before the
LLM wave.

Establishing knowledge continuity requires full-text evidence: whether
LLM-wave papers compare against credible non-LLM methods, carry forward known
limitations, test earlier empirical conclusions, and explain what the new
technology changes beyond task performance. Future work should therefore trace
\emph{intellectual lineage} alongside problem lineage. The purpose is not to
discourage revisiting established problems, but to distinguish cumulative
revisitation from rediscovery, and to prevent each technological wave from
quietly resetting the field's comparison point.

\subsection{Maturation through service, with limited interrogation}

The growth of \textsc{Service} qualifies the question posed in our title: SE
research is not \emph{just} applying LLMs. Researchers increasingly
characterize model behavior, construct evaluations, and investigate failures.
This work is necessary because conventional benchmarks do not fully represent
realistic conditions. Copilot evaluations have found
incorrect suggestions~\cite{nguyen2022copilot}; security studies have exposed
vulnerabilities and risks in developers' reliance on
assistants~\cite{pearce2022asleep,perry2023insecure}; and SWE-bench shows the
difficulty of resolving repository-level issues that require coordinated
changes across a codebase~\cite{jimenez2024swebench}. The expanding service
layer can therefore be read as a sign of technical maturation.

Yet maturation through evaluation is not equivalent to interrogation. Apply
and Service together remain the overwhelming majority of AI research, while
work questioning the paradigm's premises, problem formulations, and
alternatives remains below one in twenty AI papers. We call this portfolio-level
imbalance the \emph{interrogation gap}: the distance between the scale at which
a technological paradigm is adopted and the scale at which its assumptions,
boundaries, and consequences are examined. This is not a judgment about the
quality or intellectual depth of individual application papers. An
\textsc{Apply} contribution may embody substantial SE knowledge, and a
\textsc{Service} paper may expose failures that enable theoretical progress.
The gap concerns proportionality: evaluation expanded with AI, but
paradigm-level questioning did not.

\subsection{LLMification beyond software engineering}

SE is not alone in reorganizing research around LLMs. In CHI, LLMs appear as
research tools, system components, and simulated users, accompanied by concerns
about validity, reproducibility, and dependence on closed
models~\cite{pang2025llmification}. Reviews of engineering education, code
security, and model-driven engineering similarly describe research communities
translating a general-purpose technology into their own tasks and evaluation
practices~\cite{filippi2024engineering,basic2024security,
zhang2026modeldriven}. These studies examine LLM-selected corpora, whereas our
census measures AI research against the surrounding publication system.
Differences in venues, periods, and inclusion criteria prevent direct
cross-disciplinary rankings. A consistently designed comparative census is
needed to determine whether SE's combination of rapid growth, inherited
problems, and predominantly downstream engagement is distinctive or part of a
broader transformation of computing research.

\subsection{From blind spots to a broader research agenda}

Our results do not imply that SE should conduct less research on LLMs. Nor is
the broader agenda we propose foreign to the community. Flagship keynotes have
warned that learned systems could unsettle established SE practice and have
called for stronger assurance, neuro-symbolic foundations, human oversight, and
new accounts of correctness in AI-enabled systems
\cite{meijer2018software2,chechik2022assurance,gulwani2022ai,
shaw2026assurance}. These examples show that many ingredients of an alternative
agenda already exist in the field's agenda-setting discourse. Read alongside
their limited presence in the research-track portfolio and the survey's gap
between recognition and response, the challenge is to turn these dispersed
calls into a sustained and cumulative research agenda.

\paragraph{Build cumulatively across technological waves.}
Papers revisiting an established problem should make its pre-LLM lineage
explicit, compare against credible earlier methods when they remain relevant,
and state which prior findings the new technology preserves or overturns.
Evaluations should also distinguish improvements attributable to model
capability from improvements in SE knowledge. A higher task score demonstrates
better performance; it does not necessarily show that the field understands
the underlying problem better. Explicit lineage and stronger baselines would
make clear whether a new model extends the state of knowledge. The need for such continuity predates the LLM wave. As early as 2018,
agenda-setting discussions warned that learned systems could unsettle much of
what SE had established about development and maintenance
\cite{meijer2018software2}. 


\paragraph{Develop an evaluation and assurance science for AI-enabled
software.}
SE should turn the growing service layer into a cumulative science of when,
where, and under what conditions AI-enabled software can be trusted. This
requires moving beyond repeated comparisons of model performance toward
evaluations derived from software requirements and realistic development
contexts. Priorities include construct-valid measures, contamination and
temporal validity, longitudinal reliability, and maintenance of AI-generated
artifacts, provenance, failure recovery, accountability, and verification cost.
Calls to make assurance integral to ML-enabled systems
\cite{chechik2022assurance} and to rethink correctness, confidence, and context
under probabilistic generation~\cite{shaw2026assurance} already establish the
contours of this agenda. Repository-level benchmarks are an important step
\cite{jimenez2024swebench}, but no fixed benchmark can substitute for evidence
about systems, people, and processes after deployment. Shared evidence
standards~\cite{ralph2021empirical} should enable findings to accumulate rather than expire with each new release.

\paragraph{Develop SE-native foundations and alternatives.}
The scarcity of reframing and foundational work suggests that the field has
invested more in adapting AI to SE problems than in asking what SE contributes
that increasing model capability does not. Program semantics, execution traces,
types, specifications, architectures, dependencies, and requirements can serve
as foundations for a solution rather than auxiliary context supplied to a
model. These structures may constrain, verify, complement, or sometimes replace
probabilistic generation. The objective is not for SE researchers to compete in
training larger general-purpose models, but to develop theories and techniques
that derive their leverage from the distinctive properties of software.

The ingredients of such an agenda are already visible. Gulwani argues for
combining learned models with symbolic reasoning and programming-language
semantics~\cite{gulwani2022ai}, while Rungta demonstrates how domain-specific
languages and constraints can make correctness more tractable
\cite{rungta2025correctness}. These examples illustrate what it means for SE
structure to serve as a foundation rather than merely as a context supplied to a
general-purpose model.

\paragraph{Study SE in an AI-transformed world.}
Most current work asks how AI can perform an existing SE task. A complementary
agenda should examine how pervasive AI changes the practice of
software engineering itself. Studies of AI-assisted programming already show
that interaction, trust, and validation shape outcomes
\cite{barke2023grounded,vaithilingam2022expectation,perry2023insecure}. As
generation becomes inexpensive while verification remains costly, these
questions extend beyond individual tasks: development may reorganize around
review and assurance, while delegating implementation may alter the knowledge
developers retain, the artifacts teams require, and the allocation of
authorship and responsibility. This broader agenda is anticipated by Rinard's
call to investigate the research problems and software systems made possible by
generative AI~\cite{rinard2024generative} and Lu's argument that increasingly
autonomous generation shifts the engineering bottleneck toward verification,
decomposition, and human judgment~\cite{lu2026aiware}. Pursuing these questions
would move beyond \emph{AI for SE} toward understanding \emph{SE in an
AI-transformed world}.

Realizing this broader agenda is not solely the responsibility of individual
researchers. Venues,
reviewers, funders, datasets, and infrastructure influence which questions
appear feasible and valuable. Valuing cumulative baselines, replication,
construct validity, negative results, longitudinal evidence, and foundational
work would create room for research whose importance does not depend on the
novelty of the latest model. The opportunity is not to retreat from LLM
research, but to move from following model capabilities toward deliberately
shaping the questions, standards, and foundations through which software
engineering responds to them.



\section{Limitations and Future Work}
\label{sec:limitations}

Our census covers accepted research-track papers at ICSE, FSE, and ASE, not all
SE research; ASE 2026 was unavailable and is excluded from complete-year
comparisons. The keyword screen may miss papers using terminology outside our
lexicons, and publication data cannot establish causal displacement without
submission and acceptance data. Title--abstract coding also limits construct
resolution: one primary problem suppresses secondary contributions, stance does
not measure depth or quality, and \textsc{Established} does not demonstrate
engagement with prior knowledge. Such continuity requires full-text analysis.
Given imperfect agreement, we emphasize distributions and release paper-level
labels for audit.

FoSE respondents form a self-selected rather than representative sample. Q12's
future-oriented framing may increase AI salience, while Q9's request for one
change may suppress secondary priorities; few AI-specific proposals therefore
do not imply indifference. Survey labels came from one model coder, and the
second pass measures self-consistency rather than independent agreement. We
accordingly use the survey as qualitative resonance and its quotations as
illustrative rather than population estimates.

We exclude journals, specialized tracks, co-hosted conferences, and workshops,
which may contain work uncommon in the flagship tracks. Future work should map
whether these venues incubate ideas that later enter the main tracks or sustain
parallel research conversations.

\section{Conclusion}
The LLM wave has changed software engineering research more through scale and
redistribution than through a new problem agenda. AI became the majority of
flagship research-track output, yet $92.6\%$ of LLM-wave papers revisit
established problems, and $95.7\%$ of AI papers apply AI or evaluate and support
its use. Only $4.3\%$ interrogate the paradigm or develop alternatives. FoSE
respondents recognized corresponding risks but rarely proposed AI-specific
change. These findings do not call for less LLM research, but for turning rapid
adoption into cumulative knowledge, stronger evaluation, SE-native foundations,
and a deeper understanding of software engineering in an AI-transformed world.

\section*{Data Availability}
An anonymized replication package is available at
\url{https://figshare.com/s/faad9ea59579f14e428a} through figshare. The package contains the corpus manifest and inclusion decisions; cleaned
paper-level data and coder outputs; AI-screening lexicons; qualitative
codebooks and decision records; analysis scripts and derived tables and
figures; and the FoSE survey coding and audit materials. A README
documents the package structure and reproduction procedure. The original FoSE
responses remain available through their cited public dataset. For the
camera-ready version, we will replace the anonymous link with the archived,
version-specific DOI.

\bibliographystyle{ACM-Reference-Format}
\bibliography{ref}

\end{document}